\documentclass[aps,preprint]{revtex4}
\usepackage{amsfonts}
\usepackage{amsmath}
\usepackage{amssymb}
\usepackage{graphicx}

\input{tcilatex}

\begin{document}

\preprint{}
\title[ ]{Stability of thin-shell wormholes supported by normal matter in
Einstein-Maxwell-Gauss-Bonnet gravity}
\author{S. Habib Mazharimousavi}
\email{habib.mazhari@emu.edu.tr}
\author{M. Halilsoy}
\email{mustafa.halilsoy@emu.edu.tr}
\author{Z. Amirabi}
\email{zahra.amirabi@emu.edu.tr}
\affiliation{Department of Physics, Eastern Mediterranean University, G. Magusa, north
Cyprus, Mersin 10, Turkey.}
\pacs{PACS number}

\begin{abstract}
Recently in ( Phys. Rev. D 76, 087502 (2007) and Phys. Rev. D 77, 089903(E)
(2008)) a thin-shell wormhole has been introduced in 5-dimensional
Einstein-Maxwell-Gauss-Bonnet (EMGB) gravity which was supported by normal
matter. We wish to consider this solution and investigate its stability. Our
analysis shows that for the Gauss-Bonnet (GB) parameter $\alpha <0,$
stability regions form for a narrow band of finely-tuned mass and charge.
For the case $\alpha >0$, we iterate once more that no stable, normal matter
thin-shell wormhole exists.
\end{abstract}

\maketitle

\section{Introduction}

Whenever the agenda is about wormholes exotic matter (i.e. matter violating
the energy conditions) continues to occupy a major issue in general
relativity \cite{1}. It is a fact that Einstein's equations admit wormhole
solutions that require such matter for its maintenance. In quantum theory
temporary violation of energy conditions is permissible but in classical
physics this can hardly be justified. One way to minimize such exotic
matter, even if we can not ignore it completely, is to concentrate it on a
thin-shell. This seemed feasible, because general relativity admits such
thin-shell solutions and by employing these shells at the throat region may
provide the necessary repulsion to support the wormhole against collapse.
The ultimate aim of course, is to get rid of exotic matter completely, no
matter how small. In the 4-dimensional (4D) general relativity with a
cosmological term, however, such a dream never turned into reality. For this
reason the next search should naturally cover extensions of general
relativity to higher dimensions and with additional structures. One such
possibility that received a great deal of attention in recent years, for a
number of reasons, is the Gauss-Bonnet (GB) extension of general relativity 
\cite{2}. In the brane-world scenario our universe is modelled as a brane in
a 5D bulk universe in which the higher order curvature terms, and therefore
the GB gravity comes in naturally. Einstein-Gauss-Bonnet (EGB) gravity, with
additional sources such as Maxwell, Yang-Mills, dilaton etc. has already
been investigated extensively in the literature \cite{3}. Not to mention,
all these theories admit black hole, wormhole \cite{4} and other physically
interesting solutions. As it is the usual trend in theoretical physics, each
new parameter invokes new hopes and from that token, the GB parameter $%
\alpha $ does the same. Although the case $\alpha >0,$ has been exalted much
more than the case $\alpha <0$ in EGB gravity so far \cite{5} (and
references cited therein), it turns out here in the stable, normal matter
thin-shell wormholes that the latter comes first time to the fore.

Construction and maintenance of thin-shell wormholes has been the subject of
a large literature, so that we shall provide only a brief review here.
Instead, one class \cite{6} that made use of non-exotic matter for its
maintenance attracted our interest and we intend to analyze its stability in
this paper. This is the 5D thin-shell solution of
Einstein-Maxwell-Gauss-Bonnet (EMGB) gravity, whose radius is identified
with the minimum radius of the wormhole. For this purpose we employ radial,
linear perturbations to cast the motion into a potential-well problem in the
background. In doing this, a reasonable assumption employed routinely, which
is adopted here also, is to relate pressure and energy density by a linear
expression \cite{7}. For special choices of parameters we obtain islands of
stability for such wormholes. To this end, we make use of numerical
computation and plotting since the problem involves highly intricate
functions for an analytical treatment.

The paper is organized as follows. In Sec. II the 5D EMGB thin-shell
wormhole formalism has been reviewed briefly. We perturb the wormhole
through radial linear perturbation and cast the problem into a
potential-well problem in Sec. III. In Sec. IV we impose constraint
conditions on parameters to determine possible stable regions through
numerical analysis. The paper ends with Conclusion which appears in Sec. V.

\section{A brief review of 5D EMGB thin-shells}

The action of EMGB gravity in 5D (without cosmological constant, i.e. $%
\Lambda =0$) is%
\begin{equation}
S=\kappa \int \sqrt{\left\vert g\right\vert }d^{5}x\left( R+\alpha \mathcal{L%
}_{GB}-\frac{1}{4}F_{\mu \nu }F^{\mu \nu }\right)
\end{equation}%
in which $\kappa $ is related to the 5D Newton constant and $\alpha $ is the
GB parameter. Beside the Maxwell Lagrangian the GB Lagrangian $\mathcal{L}%
_{GB}$ consists of the quadratic scalar invariants in the combination 
\begin{equation}
\mathcal{L}_{GB}=R^{2}-4R_{\mu \nu }R^{\mu \nu }+R_{\mu \nu \rho \sigma
}R^{\mu \nu \rho \sigma }
\end{equation}%
in which $R=$scalar curvature, $R_{\mu \nu }=$Ricci tensor and $R_{\mu \nu
\rho \sigma }=$Riemann tensor. Variational principle of $S$ with respect to $%
g_{\mu \nu }$ yields 
\begin{equation}
G_{\mu \nu }+2\alpha H_{\mu \nu }^{\ }=\kappa ^{2}T_{\mu \nu }
\end{equation}%
where the Lovelock ($H_{\mu \nu }$) and Maxwell ($T_{\mu \nu }$) tensors
respectively are 
\begin{equation}
H_{\mu \nu }^{\ }=2(-R_{\mu \text{ \ \ }}^{\ \sigma \kappa \tau }R_{\nu
\sigma \kappa \tau }-2R_{\ \mu \rho \nu \sigma }^{\quad }R^{\rho \sigma
}-2R_{\mu \sigma }R_{\text{ \ }\nu }^{\sigma }+RR_{\mu \nu }^{\ })-\frac{1}{2%
}g_{\mu \nu }^{\ }\tciLaplace _{GB},
\end{equation}

\begin{equation}
T_{\mu \nu }=F_{\mu \alpha }F_{\nu }^{\alpha }-\frac{1}{4}g_{\mu \nu
}F_{\alpha \beta }F^{\alpha \beta }.
\end{equation}%
The Einstein tensor $G_{\mu \nu }$ is to be found from our metric ansatz%
\begin{equation}
ds^{2}=-f\left( r\right) dt^{2}+\frac{dr^{2}}{f\left( r\right) }+r^{2}\left(
d\theta ^{2}+\sin ^{2}\theta \left( d\phi ^{2}+\sin ^{2}\phi d\psi
^{2}\right) \right) ,
\end{equation}%
in which $f\left( r\right) $ will be determined from (3). A thin-shell
wormhole is constructed in EMGB theory as follows. Two copies of the
spacetime are chosen from which the regions%
\begin{equation}
M_{1,2}=\left\{ r_{1,2}\leq a,\text{ \ }a>r_{h}\right\} 
\end{equation}%
are removed. We note that $a$ will be identified in the sequel as the radius
of the thin-shell and $r_{h}$ stands for the event horizon radius. (Note
that our notation $a$ corresponds to $b$ in Ref. \cite{6}. Other notations
all agree with those in Ref. \cite{6}). The boundary, time-like surface $%
\Sigma _{1,2\text{ }}$of each $M_{1,2}$, accordingly will be%
\begin{equation}
\Sigma _{1,2\text{ }}=\left\{ r_{1,2}=a,\text{ \ }a>r_{h}\right\} .
\end{equation}%
Next, these surfaces are identified on $r=a$ with a surface energy-momentum
of a thin-shell such that geodesic completeness holds. Following the
Darmois-Israel formalism \cite{8} in terms of the original coordinates $%
x^{\gamma }=\left( t,r,\theta ,\phi ,\psi \right) ,$ we define $\xi
^{a}=\left( \tau ,\theta ,\phi ,\psi \right) $, with $\tau $ the proper
time. The GB extension of the thin-shell EM theory requires further
modifications. This entails the generalized Darmois-Israel boundary
conditions \cite{9}, where the surface energy-momentum tensor is expressed
by $S_{a}^{b}=$diag$\left( \sigma ,p_{\theta },p_{\phi },p_{\psi }\right) $.
We are interested in the thin-shell geometry whose radius is assumed a
function of $\tau $, so that the hypersurface becomes%
\begin{equation}
\Sigma :f\left( r,\tau \right) =r-a\left( \tau \right) =0.
\end{equation}%
The generalized Darmois-Israel conditions on $\Sigma $ take the form 
\begin{equation}
2\left\langle K_{ab}-Kh_{ab}\right\rangle +4\alpha \left\langle
3J_{ab}-Jh_{ab}+2P_{acdb}K^{cd}\right\rangle =-\kappa ^{2}S_{ab},
\end{equation}%
where a bracket implies a jump across $\Sigma $, and $%
h_{ab}=g_{ab}-n_{a}n_{b}$ is the induced metric on $\Sigma $ with normal
vector $n_{a}.$ $K_{ab}$ is the extrinsic curvature (with trace $K$),
defined by 
\begin{equation}
K_{ab}^{\pm }=-n_{c}^{\pm }\left( \frac{\partial ^{2}x^{c}}{\partial \xi
^{a}\partial \xi ^{b}}+\Gamma _{mn}^{c}\frac{\partial x^{m}}{\partial \xi
^{a}}\frac{\partial x^{n}}{\partial \xi ^{b}}\right) _{r=a}.
\end{equation}%
The remaining expressions are as follows. The divergence-free part of the
Riemann tensor $P_{abcd}$ and the tensor $J_{ab}$ (with trace $J$) are given
by%
\begin{eqnarray}
P_{abcd} &=&R_{abcd}+\left( R_{bc}h_{da}-R_{bd}h_{ca}\right) -\left(
R_{ac}h_{db}-R_{ad}h_{cb}\right) +\frac{1}{2}R\left(
h_{ac}h_{db}-h_{ad}h_{cb}\right) , \\
J_{ab} &=&\frac{1}{3}\left[
2KK_{ac}K_{b}^{c}+K_{cd}K^{cd}K_{ab}-2K_{ac}K^{cd}K_{ab}-K^{2}K_{ab}\right] .
\end{eqnarray}%
The EMGB solution that will be employed as a thin-shell solution with a
normal matter \cite{6} is given by (with $\Lambda =0$)%
\begin{equation}
f\left( r\right) =1+\frac{r^{2}}{4\alpha }\left( 1-\sqrt{1+\frac{8\alpha }{%
r^{4}}\left( \frac{2M}{\pi }-\frac{Q^{2}}{3r^{2}}\right) }\right) 
\end{equation}%
with constants, $M=$mass and $Q=$charge. For a black hole solution the inner
($r_{-}$) and event horizons ($r_{+}=r_{h}$) are 
\begin{equation}
r_{\pm }=\sqrt{\frac{M}{\pi }-\alpha \pm \left[ \left( \frac{M}{\pi }-\alpha
\right) ^{2}-\frac{Q^{2}}{3}\right] ^{1/2}}.
\end{equation}%
By employing the solution (14) we determine the surface energy-momentum on
the thin-shell, which will play the major role in the perturbation. We shall
address this problem in the next section.

\section{Radial, Linear perturbation of the thin-shell wormhole with normal
matter}

In order to study the radial perturbations of the wormhole we take the
throat radius as a function of the proper time, i.e., $a=a\left( \tau
\right) $. Based on the generalized Birkhoff theorem, for $r>a\left( \tau
\right) $ the geometry will be given still by (6). For the metric function $%
f\left( r\right) $ given in (14) one finds the energy density and pressures
as \cite{6} 
\begin{eqnarray}
\sigma  &=&-S_{\tau }^{\tau }=-\frac{\Delta }{4\pi }\left[ \frac{3}{a}-\frac{%
4\alpha }{a^{3}}\left( \Delta ^{2}-3\left( 1+\dot{a}^{2}\right) \right) %
\right] , \\
S_{\theta }^{\theta } &=&S_{\phi }^{\phi }=S_{\psi }^{\psi }=p=\frac{1}{4\pi 
}\left[ \frac{2\Delta }{a}+\frac{\ell }{\Delta }-\frac{4\alpha }{a^{2}}%
\left( \ell \Delta -\frac{\ell }{\Delta }\left( 1+\dot{a}^{2}\right) -2\ddot{%
a}\Delta \right) \right] ,
\end{eqnarray}%
where $\ell =\ddot{a}+f^{\prime }\left( a\right) /2$ and $\Delta =\sqrt{%
f\left( a\right) +\dot{a}^{2}}$ in which 
\begin{equation}
f\left( a\right) =1+\frac{a^{2}}{4\alpha }\left( 1-\sqrt{1+\frac{8\alpha }{%
a^{4}}\left( \frac{2M}{\pi }-\frac{Q^{2}}{3a^{2}}\right) }\right) .
\end{equation}%
We note that in our notation a 'dot' denotes derivative with respect to the
proper time $\tau $ and a 'prime' implies differentiation with respect to
the argument of the function. By a simple substitution one can show that,
the conservation equation 
\begin{equation}
\frac{d}{d\tau }\left( \sigma a^{3}\right) +p\frac{d}{d\tau }\left(
a^{3}\right) =0.
\end{equation}%
is satisfied. The static configuration of radius $a_{0}$ has the following
density and pressures 
\begin{eqnarray}
\sigma _{0} &=&-\frac{\sqrt{f\left( a_{0}\right) }}{4\pi }\left[ \frac{3}{%
a_{0}}-\frac{4\alpha }{a_{0}^{3}}\left( f\left( a_{0}\right) -3\right) %
\right] , \\
p_{0} &=&\frac{\sqrt{f\left( a_{0}\right) }}{4\pi }\left[ \frac{2}{a_{0}}+%
\frac{f^{\prime }\left( a_{0}\right) }{2f\left( a_{0}\right) }-\frac{2\alpha 
}{a_{0}^{2}}\frac{f^{\prime }\left( a_{0}\right) }{f\left( a_{0}\right) }%
\left( f\left( a_{0}\right) -1\right) \right] .
\end{eqnarray}

In what follows we shall study small radial perturbations around the radius
of equilibrium $a_{0}.$ To this end we adapt a linear relation between $p$
and $\sigma $ as \cite{7} 
\begin{equation}
p=p_{0}+\beta ^{2}\left( \sigma -\sigma _{0}\right) .
\end{equation}%
Here since we are interested in the wormholes which are supported by normal
matter, $\beta ^{2}$ is the speed of sound. By virtue of Eq.s (19) and (22)
we find the energy density in the form 
\begin{equation}
\sigma \left( a\right) =\left( \frac{\sigma _{0+}p_{0}}{\beta ^{2}+1}\right)
\left( \frac{a_{0}}{a}\right) ^{3\left( \beta ^{2}+1\right) }+\frac{\beta
^{2}\sigma _{0-}p_{0}}{\beta ^{2}+1}.
\end{equation}%
This, together with (16) lead us to the equation of motion for the radius of
the throat, which reads%
\begin{equation}
-\frac{\sqrt{f\left( a\right) +\dot{a}^{2}}}{4\pi }\left[ \frac{3}{a}-\frac{%
4\alpha }{a^{3}}\left( f\left( a\right) -3-2\dot{a}^{2}\right) \right]
=\left( \frac{\sigma _{0+}p_{0}}{\beta ^{2}+1}\right) \left( \frac{a_{0}}{a}%
\right) ^{3\left( \beta ^{2}+1\right) }+\frac{\beta ^{2}\sigma _{0-}p_{0}}{%
\beta ^{2}+1}.
\end{equation}%
After some manipulation this can be cast into 
\begin{equation}
\dot{a}^{2}+V\left( a\right) =0,
\end{equation}%
where 
\begin{equation}
V\left( a\right) =f\left( a\right) -\left( \left[ \sqrt{A^{2}+B^{3}}-A\right]
^{1/3}-\frac{B}{\left[ \sqrt{A^{2}+B^{3}}-A\right] ^{1/3}}\right) ^{2}
\end{equation}%
in which the functions $A$ and $B$ are 
\begin{eqnarray}
A &=&\frac{\pi a^{3}}{4\alpha }\left[ \left( \frac{\sigma _{0+}p_{0}}{\beta
^{2}+1}\right) \left( \frac{a_{0}}{a}\right) ^{3\left( \beta ^{2}+1\right) }+%
\frac{\beta ^{2}\sigma _{0-}p_{0}}{\beta ^{2}+1}\right] , \\
B &=&\frac{a^{2}}{8\alpha }+\frac{1-f\left( a\right) }{2}.
\end{eqnarray}%
We notice that $V\left( a\right) ,$ and more tediously $V^{\prime }\left(
a\right) ,$ both vanish at $a=a_{0}.$ The stability requirement for
equilibrium reduces therefore to the determination of $\ V^{\prime \prime
}(a_{0})>0,$ and it is needless to add that, $V\left( a\right) $ is
complicated enough for an immediate analytical result. For this reason we
shall proceed through numerical calculation to see whether stability
regions/ islands develop or not. Since the hopes for obtaining thin-shell
wormholes with normal matter when $\alpha >0,$ have already been dashed \cite%
{5}, we shall investigate here only the case for $\alpha <0.$

In order to analyze the behavior of $V\left( a\right) $ (and its second
derivative) we introduce new parameterization as follows%
\begin{equation}
\tilde{a}^{2}=-\frac{a^{2}}{\alpha },\text{ }m=-\frac{16M}{\pi \alpha },%
\text{ }q^{2}=\frac{8Q^{2}}{3\alpha ^{2}},\text{ }\tilde{\sigma}_{0}=\sqrt{%
-\alpha }\sigma _{0},\text{ }p_{0}=\sqrt{-\alpha }p_{0}
\end{equation}%
Accordingly, our new variables $f\left( \tilde{a}\right) ,$ $\tilde{\sigma}%
_{0},$ $\tilde{p}_{0},$ $A$ and $B$ take the following forms 
\begin{equation}
f\left( \tilde{a}\right) =1-\frac{\tilde{a}^{2}}{4}+\frac{\tilde{a}^{2}}{4}%
\sqrt{1-\frac{m}{\tilde{a}^{4}}+\frac{q^{2}}{\tilde{a}^{6}}}
\end{equation}%
and 
\begin{eqnarray}
\tilde{\sigma}_{0} &=&-\frac{\sqrt{f\left( \tilde{a}_{0}\right) }}{4\pi }%
\left[ \frac{3}{\tilde{a}_{0}}+\frac{4}{\tilde{a}_{0}^{3}}\left( f\left( 
\tilde{a}_{0}\right) -3\right) \right] , \\
\tilde{p}_{0} &=&\frac{\sqrt{f\left( \tilde{a}_{0}\right) }}{4\pi }\left[ 
\frac{2}{\tilde{a}_{0}}+\frac{f^{\prime }\left( \tilde{a}_{0}\right) }{%
2f\left( \tilde{a}_{0}\right) }+\frac{2}{\tilde{a}_{0}^{2}}\frac{f^{\prime
}\left( \tilde{a}_{0}\right) }{f\left( \tilde{a}_{0}\right) }\left( f\left( 
\tilde{a}_{0}\right) -1\right) \right] ,
\end{eqnarray}%
\begin{eqnarray}
A &=&-\frac{\pi \tilde{a}^{3}}{4}\left[ \left( \frac{\tilde{\sigma}_{0+}%
\tilde{p}_{0}}{\beta ^{2}+1}\right) \left( \frac{\tilde{a}_{0}}{\tilde{a}}%
\right) ^{3\left( \beta ^{2}+1\right) }+\frac{\beta ^{2}\tilde{\sigma}_{0-}%
\tilde{p}_{0}}{\beta ^{2}+1}\right] , \\
B &=&-\frac{\tilde{a}^{2}}{8}+\frac{1-f\left( \tilde{a}\right) }{2}.
\end{eqnarray}%
Following this parametrization our Eq. (25) takes the form 
\begin{equation}
\left( \frac{d\tilde{a}}{d\tau }\right) ^{2}+\tilde{V}\left( \tilde{a}%
\right) =0,
\end{equation}%
where%
\begin{equation}
\tilde{V}\left( \tilde{a}\right) =-\frac{V\left( \tilde{a}\right) }{\alpha }.
\end{equation}%
In the next section we explore all possible constraints on our parameters
that must satisfy to materialize a stable, normal matter wormhole through
the requirement $V^{\prime \prime }\left( \tilde{a}\right) >0.$

\section{Constraints versus finely-tuned parameters and second derivative
plots of the potential}

$i$) Starting from the metric function we must have

\begin{equation}
1-\frac{m}{\tilde{a}_{0}^{4}}+\frac{q^{2}}{\tilde{a}_{0}^{6}}\geq 0.
\end{equation}

$ii$) In the potential, the reality condition requires also that 
\begin{equation}
A^{2}+B^{3}\geq 0.
\end{equation}%
At the location of the throat this amounts to 
\begin{equation}
\left( -\frac{\pi \tilde{a}_{0}^{3}}{4}\tilde{\sigma}_{0}\right) ^{2}+\left(
-\frac{\tilde{a}_{0}^{2}}{8}+\frac{1-f\left( \tilde{a}_{0}\right) }{2}%
\right) ^{3}\geq 0
\end{equation}%
or after some manipulation it yields 
\begin{equation}
f\left( \tilde{a}_{0}\right) -2+\frac{\tilde{a}_{0}^{2}}{2}\leq 0.
\end{equation}%
This is equivalent to%
\begin{equation}
0\leq 1-\frac{m}{\tilde{a}_{0}^{4}}+\frac{q^{2}}{\tilde{a}_{0}^{6}}\leq
\left( \frac{4}{\tilde{a}_{0}^{2}}-1\right) ^{2}.
\end{equation}

$iii$) Our last constraint condition concerns, the positivity of the energy
density, which means that 
\begin{equation}
\tilde{\sigma}_{0}>0.
\end{equation}%
This implies, from (31) that 
\begin{equation}
\left[ \frac{3}{\tilde{a}_{0}}+\frac{4}{\tilde{a}_{0}^{3}}\left( f\left( 
\tilde{a}_{0}\right) -3\right) \right] <0
\end{equation}%
or equivalently%
\begin{equation}
0\leq 1-\frac{m}{\tilde{a}_{0}^{4}}+\frac{q^{2}}{\tilde{a}_{0}^{6}}<4\left( 
\frac{4}{\tilde{a}_{0}^{2}}-1\right) ^{2}.
\end{equation}%
It is remarkable to observe now that the foregoing constraints ($i-iii$) on
our parameters can all be expressed as a single constraint condition, namely 
\begin{equation}
0\leq 1-\frac{m}{\tilde{a}_{0}^{4}}+\frac{q^{2}}{\tilde{a}_{0}^{6}}\leq
\left( \frac{4}{\tilde{a}_{0}^{2}}-1\right) ^{2}.
\end{equation}

We plot $\tilde{V}^{\prime \prime }\left( \tilde{a}\right) $ from (26) for
various fixed values of mass and charge, as a projection into the plane with
coordinates $\beta $ and $\tilde{a}_{0}.$ In other words, we search and
identify the regions for which $\tilde{V}^{\prime \prime }\left( \tilde{a}%
\right) >0$, in $3-$dimensional figures considered as a projection in the $%
\left( \beta ,\tilde{a}_{0}\right) $ plane. The metric function $f\left(
r\right) $ and energy density $\tilde{\sigma}_{0}>0,$ behavior also are
given in Fig.s 1-4. It is evident from Fig.s 1-4 that for increasing charge
the stability regions shrink to smaller domains and tends ultimately to
disappear completely. For smaller $\tilde{a}_{0}$ bounds\ we obtain
fluctuations in $\tilde{V}^{\prime \prime }\left( \tilde{a}\right) ,$ which
is smooth otherwise.

In each plot it is observed that the maximum of $\tilde{V}^{\prime \prime
}\left( \tilde{a}\right) $ occurs at the right-below corner (say, at $%
a_{\max }$) \ which decreases to the left (with $\tilde{a}_{0}$) and in the
upward direction (with $\beta $). Beyond certain limit (say $a_{\min }$),
the region of instability takes the start. The proper time domain of
stability can be computed from (35) as 
\begin{equation}
\Delta \tau =\int_{a_{\min }}^{a_{\max }}\frac{d\tilde{a}}{\sqrt{-V\left( 
\tilde{a}\right) }}.
\end{equation}%
From a distant observer's point of view the timespan $\Delta t$ can be found
by using the radial geodesics Lagrangian which admits the energy integral 
\begin{equation}
f\left( \frac{dt}{d\tau }\right) =E_{\circ }=const.
\end{equation}%
This gives the lifetime of each stability region determined by 
\begin{equation}
\Delta t=\frac{1}{E_{\circ }}\int_{a_{\min }}^{a_{\max }}\frac{d\tilde{a}}{%
f\left( \tilde{a}\right) \sqrt{-V\left( \tilde{a}\right) }}.
\end{equation}%
Once $a_{\min }$ ($a_{\max }$) are found numerically, assuming that no zeros
of $f\left( \tilde{a}\right) $ and $V\left( \tilde{a}\right) $ occurs for $%
a_{\min }<a<a_{\max },$ the lifespan of each stability island can be
determined. We must admit that the mathematical complexity discouraged us to
search for possible metastable region that may be triggered by employing a
semi-classical treatment.

\section{Conclusion}

Our numerical analysis shows that for $\alpha <0,$ and specific ranges of
mass and charge the 5D EMGB thin-shell wormholes with normal matter can be
made stable against linear, radial perturbations. The fact that for $\alpha
>0$ there is no such wormholes is well-known. The magnitude of $\alpha $ is
irrelevant to the stability analysis. This reflects the universality of
wormholes in parallel with black holes, i.e., the fact that they arise at
each scale. Stable regions develop for each set of finely-tuned parameters
which determine the lifespan of each such region. Beyond those regions
instability takes the start. Our study concerns entirely the exact EMGB
gravity solution given in Ref. \cite{6}. It is our belief that beside EMGB
theory in different theories also such stable, normal-matter wormholes are
abound, which will be our next venture in this line of research.

\bigskip

\bigskip \textbf{Figure Captions:}

Fig. 1: $\tilde{V}^{\prime \prime }\left( \tilde{a}\right) >0$ region ($%
m=0.5 $, $q=1.0$) for various ranges of $\beta $ and $\tilde{a}_{0}.$ The
lower and upper limits of the parameters are evident in the figure. The
metric function $f\left( \tilde{r}\right) $ and $\tilde{\sigma}_{0}>0$, are
also indicated in the smaller figures.

Fig. 2: $\tilde{V}^{\prime \prime }\left( \tilde{a}\right) >0$ plot for $%
m=1.0$, $q=1.5$. The stability region is seen clearly to shrink with the
increasing charge. This effect reflects also to the $\tilde{\sigma}_{0}>0,$
behavior.

Fig. 3: The stability region for $m=1.0$, $q=2.0$, is seen to shift outward
and get smaller.

Fig. 4: For fixed mass $m=1.0$ but increased charge $q=2.5$ it is clearly
seen that the stability region and the associated energy density both get
further reduced.

\end{document}